\documentclass[%
reprint,
superscriptaddress,
amsmath,amssymb,
aps,
]{revtex4-2}
\usepackage{graphicx}
\usepackage{dcolumn}
\usepackage{bm}
\usepackage{xcolor}
\usepackage{siunitx}

\begin{document}

\title{Kerr Nonlinearity Induced Nonreciprocity in dissipatively coupled resonators}

\author{Qingtian Miao}
\email{qm8@tamu.edu}
  \affiliation{Department of Physics and Astronomy, Texas A\&M University, Texas 77843, USA}
\author{G.S. Agarwal}
\email{Girish.Agarwal@ag.tamu.edu}
 \affiliation{Institute for Quantum Science and Engineering, Department of Biological and Agricultural Engineering, Department of Physics and Astronomy, Texas A\&M University, College Station, Texas 77843, USA}

\date{\today}

\begin{abstract}
Nonlinearity induced nonreciprocity is studied in a system comprising two resonators coupled to a one-dimensional waveguide when the linear system does not exhibit nonreciprocity. The analysis is based on the Hamiltonian of the coupled system and includes the dissipative coupling between the waveguide and resonators, along with the input-output relations. We consider a large number of scenarios which can lead to nonreciprocity. We pay special attention to the case when the linear system does not exhibit nonreciprocal behavior. In this case, we show how very significant nonreciprocal behavior can result from Kerr nonlinearities. We find that the bistability of the nonlinear system can aid in achieving large nonreciprocity. Additionally, We bring out nonreciprocity in the excitation of each resonator, which can be monitored independently. Our results highlight the profound influence of nonlinearity on nonreciprocal behavior, offering a new avenue for controlling light propagation in integrated photonic circuits. Nonlinearity induced nonreciprocity would lead to significant nonreciprocity in quantum fluctuations when our system is treated quantum mechanically.
\end{abstract}

\maketitle

\section{Introduction}

The nonreciprocal propagation of light is increasingly attracting attention \cite{doyeux2017giant,caloz2018electromagnetic,sohn2018time,kim2017complete,mahmoud2015all}. Time reversal symmetry breaking is a necessary condition for nonreciprocity. Historically, this has been achieved with magneto-optical (Faraday-rotation) crystals that require an external magnetic bias. However, these crystals are not compatible with semiconductor chip integration due to their bulkiness and complexity. More generally, nonreciprocity often naturally arises if the medium is chiral as is typical in magnetic systems \cite{wang2019nonreciprocity}. Recent studies in chiral quantum optics have shown that strong light confinement in certain structures can lock the local light polarization along its direction of travel, yielding direction-dependent emission characteristics \cite{lodahl2017chiral}. It is more challenging to produce nonreciprocity in achiral systems. Some standard demonstrations of nonreciprocity use phase matching conditions in a nonlinear medium \cite{dong2015brillouin,kim2015non}. Clearly, the phase matching conditions can not be simultaneously satisfied for two opposite directions of propagation and this would result in nonreciprocity. A recent demonstration of nonreciprocity is based on the intrinsic nonlinearities of two-level superconducting artificial atoms \cite{hamann2018nonreciprocity}. Additionally, the use of synthetic electric and magnetic fields, or a suitable modulation of the refractive index, are attracting considerable attention for producing nonreciprocity without the need for traditional magnetic fields \cite{sounas2017non,estep2014magnetic,peterson2019strong,biehs2023enhancement,shaltout2015time,li2018floquet,koutserimpas2018nonreciprocal,fang2012realizing}. Synthetic fields are preferable over magnetic fields \cite{reiskarimian2016magnetic,qin2014nonreciprocal} as implementation of synthetic fields is straightforward. It should be noted that the nonreciprocal transport has consequences in several other contexts like heat transport \cite{ott2020thermal,khandekar2019thermal,khandekar2015radiative,rodriguez2024heat}, one-way amplification \cite{zou2024dissipative}; quantum fluctuations and quantum entanglement \cite{tang2022quantum,graf2022nonreciprocity,lu2021nonreciprocity,wanjura2023quadrature,agarwal2012spontaneous}. One would like to have systems that are scalable and integrable leading to the theoretical development and experimental realization of nonreciprocal devices such as isolators, circulators, and directional amplifiers. 

Waveguide-coupled resonator systems are increasingly capturing the interest of researchers for their pivotal role in facilitating nonreciprocal light propagation, a critical mechanism for unidirectional wave transport, and a foundational element for future quantum-information networks. These systems are especially attractive because of the waveguide-mediated dissipative coupling and the resulting bound states in continuum which can be instrumental in enhancing nonreciprocal effects \cite{hamann2018nonreciprocity,zou2024dissipative,okawachi2023chip,nair2021enhanced,zhang2020breaking}. One way to produce nonreciprocity in such systems is to use both dissipative and dispersive couplings among resonators. Unidirectional propagation was demonstrated using the combined effect of the dispersive and dissipative couplings \cite{wang2019nonreciprocity}. The question that we address in this work is how to achieve nonreciprocity if the resonators are far apart so that dispersive coupling is almost zero. One way is to use Kerr nonlinearity and unidirectional propagation of waves inside the resonators as would be the case of resonators coupled by a waveguide \cite{roy2010few,cotrufo2021nonlinearity1,cotrufo2021nonlinearity2}. This is different from the bidirectional propagation inside the resonator, which always produces nonreciprocity as demonstrated in several experiments \cite{del2018microresonator,del2017symmetry}. Generally, Kerr nonlinearity has been small for most materials and requires higher powers however with the development of superconducting circuits, this is no longer the case as the Kerr nonlinearity is then fairly large \cite{zoepfl2023kerr}.

The manuscript is organized as follows. In Section \ref{MB} we present details of our model and use semiclassical approach and temporal coupled-mode theory (TCMT) \cite{suh2004temporal,zhao2019connection} to obtain basic equations for the fields in the two resonators dissipatively coupled to a waveguide. We also present input-output relations. In Section \ref{trans} we derive equations for the transmission amplitudes which depend nonlinearly on the fields in the nonlinear medium. In Section \ref{linear}, we derive conditions when nonreciprocity without nonlinearity is possible and establish connection with some of the existing results. In Section \ref{nonlinear}, we conduct a thorough analysis to understand how the system’s parameters affect nonreciprocal behavior, highlighting the substantial nonreciprocity due to nonlinearity. In certain parameter domain the system exhibits bistability and then we present nonreciprocity in bistable transmission. In Section \ref{nex} we give nonreciprocity in the excitation of the resonators. In Section \ref{appen} we discuss the popular system consisting of a magnetic system coupled dissipatively to a resonator \cite{yang2020unconventional,wang2019nonreciprocity,pan2022bistability}. Our insights not only enhance the comprehension of nonreciprocal behavior in both linear and nonlinear systems but also facilitate the development of novel integrated photonic circuit designs, where the control of light propagation is crucial. We finally conclude with remarks about nonreciprocity in quantum fluctuations if the system is described by full quantum theory.

\section{Model and Basic Equations}
\label{MB}

\begin{figure}[b]
	\includegraphics[width=8.6cm]{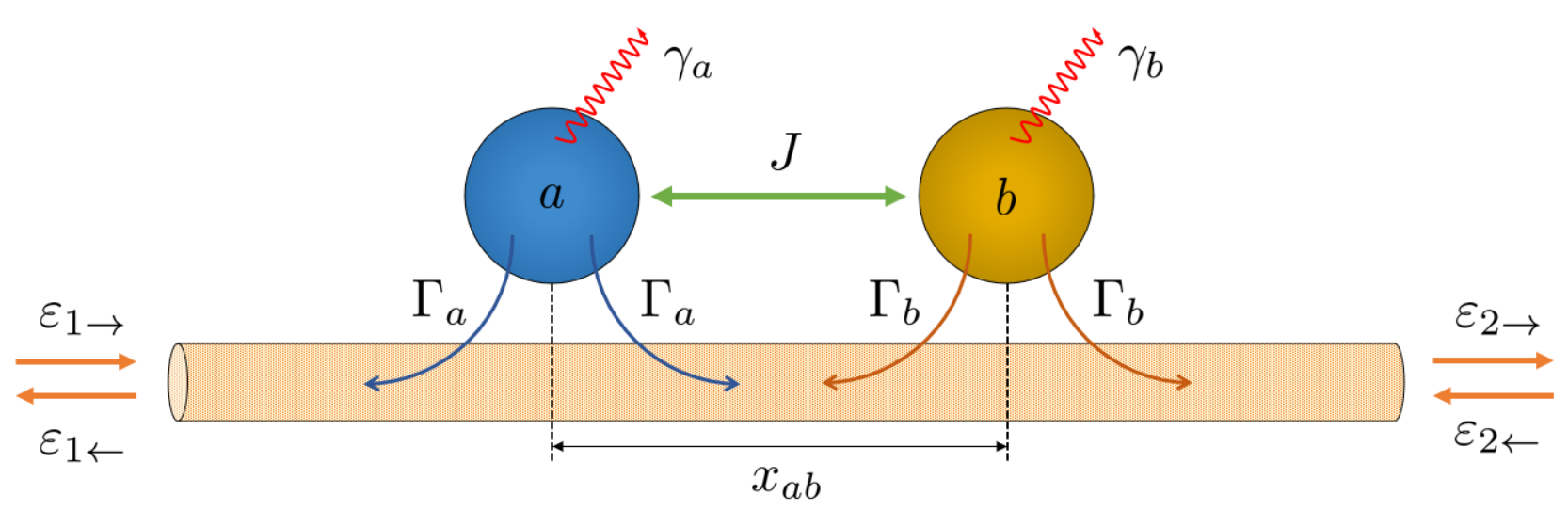}
	\caption{Two resonators $a$ and $b$, coupled through a parameter $J$, are connected to a one-dimensional waveguide, facilitating the interactions with incoming and outgoing waves. The separation $x_{ab}$ between resonators leads to a phase shift in wave propagation. $\Gamma_{a,b}$ denotes the external damping rates, and $\gamma_{a,b}$ denotes the intrinsic damping rates.}
	\label{model}
\end{figure}
We consider a system composed of two resonators $a$ and $b$ coupled to a one-dimensional waveguide, as depicted in Fig. \ref{model}. The resonators have resonance frequencies $\omega_a$ and $\omega_b$, respectively. Resonator $a$ can exhibit Kerr nonlinearity. Positioned at a distance $x_{ab}$ apart, the two resonators are coupled directly via a complex parameter $J$. In addition, they also interact through the propagating waves in the waveguide. The waveguide modes, both incoming ($\varepsilon_{1 \rightarrow}$, $\varepsilon_{2 \leftarrow}$) and outgoing ($\varepsilon_{1 \leftarrow}$, $\varepsilon_{2 \rightarrow}$), where $1,2$ specifies the port, are coherent and operating at the driving frequency $\omega_d$. The resonant modes are excited by the incoming waves ($\varepsilon_{1 \rightarrow}$, $\varepsilon_{2 \leftarrow}$) of two ports with some coupling constants $k_{a\ell}$, $k_{b\ell}$ ($\ell=1,2$). We can write the Hamiltonian of the system (we set $\hbar=1$) in the rotating frame of drive field as
\begin{equation}
	\begin{aligned}
		H&=\Delta_aa^\dagger a+\Delta_b b^\dagger b+Ja^\dagger b+J^*b^\dagger a+Ua^{\dagger2}a^2\\
		&\!\!\!\!\!\!\!\!\!\!+i( k_{a1}\varepsilon_{1\rightarrow} a^\dagger-k_{a1}^* a\varepsilon_{1\rightarrow}^\dagger)+i(k_{a2}e^{i\phi}\varepsilon_{2\leftarrow}  a^\dagger -k_{a2}^*e^{-i\phi}  a\varepsilon_{2\leftarrow}^\dagger)\\
		&\!\!\!\!\!\!\!\!\!\!+i(k_{b1}e^{i\phi}\varepsilon_{1\rightarrow} b^\dagger-k_{b1}^*e^{-i\phi} b\varepsilon_{1\rightarrow}^\dagger)+i(k_{b2}\varepsilon_{2\leftarrow} b^\dagger-k_{b2}^* b\varepsilon_{2\leftarrow}^\dagger).
	\end{aligned}
	\label{ha1}
\end{equation}
Here, $\Delta_a=\omega_a-\omega_d$, $\Delta_b=\omega_b-\omega_d$ are the detuning of the resonant frequencies $\omega_a$ and $\omega_b$ compared to the driving frequency, $\omega_d$. $U$ quantifies the strength of third-order nonlinearity in mode $a$. The phase shift is $\phi=\omega_d x_{ab}/v_p$, experienced by the drive as it travels between the resonators, with $v_p$ denoting the phase velocity of the drive. 

In what follows, we use the semiclassical description and adopt the TCMT. The dynamics of the system in the rotating frame of the drive can be given as 
\begin{equation}
	\begin{aligned}
		\frac{d}{dt}\mathbf c&=-i\mathbf\Delta\mathbf c-\mathbf \Gamma\mathbf c+\mathbf K\pmb \varepsilon_{\rm in}\\
		&\!\!\!\!\!\!-\left(\begin{array}{cc}
			0 & iJ \\
			iJ^* & 0
		\end{array}
		\right)\mathbf c-2iU(a^\dagger a)\left(\begin{array}{cc}
			1 & 0\\
			0 & 0
		\end{array}
		\right)\mathbf c,
	\end{aligned}
	\label{tcmt11}
\end{equation}
\begin{equation}
	\pmb \varepsilon_{\rm out}=\mathbf C\left(\pmb \varepsilon_{\rm in}-\mathbf D^\dagger\mathbf c\right),
	\label{tcmt12}
\end{equation}
where $\mathbf c=(a,b)^T$, $\pmb \varepsilon_{\rm in}=(\varepsilon_{1\rightarrow},\varepsilon_{2\leftarrow})^T$, $\pmb \varepsilon_{\rm out}=(\varepsilon_{1\leftarrow},\varepsilon_{2\rightarrow})^T$, $\mathbf \Delta=\left(\begin{array}{cc}
	\Delta_a-i\gamma_a & 0 \\
	0 & \Delta_b-i\gamma_b 
\end{array}
\right)$ with the intrinsic damping rates $\gamma_a$, $\gamma_b$ due to the material loss included, $\mathbf C=\left(\begin{array}{cc}
	0 & e^{i\phi} \\
	e^{i\phi} & 0
\end{array}
\right)$ for the direct scattering in the absence of resonators. The matrix $\mathbf \Gamma$ characterizes the exponential decay process. The matrices $\mathbf K$ and $\mathbf D^\dagger$ describe the coupling of the resonator to the input and output fields, respectively. The coupling matrix $\mathbf K$ are followed based on the structure of the Hamiltonian [Eq. (\ref{ha1})], 
\begin{equation}
	\mathbf K=\left(\begin{array}{cc}
		k_{a1} & k_{a2}e^{i\phi} \\
		k_{b1}e^{i\phi} & k_{b2}
	\end{array}
	\right).
\end{equation} 
In order to connect the $\mathbf\Gamma$ and $\mathbf D$ matrices to the coupling parameters $k_{a\ell}$ and $k_{b\ell}$, we look into the energy conservation constraint when $\gamma_a=\gamma_b=J=U=0$. Followed by energy conservation, $d(\mathbf c^\dagger\mathbf c)/dt=\pmb \varepsilon_{\rm in}^\dagger\pmb \varepsilon_{\rm in}-\pmb \varepsilon_{\rm out}^\dagger\pmb \varepsilon_{\rm out}$. With an assumption of no incoming waves, $\pmb \varepsilon_{\rm in}=0$, the rate of change of energy for resonators becomes $d(\mathbf c^\dagger\mathbf c)/dt=d(a^\dagger a+b^\dagger b)/dt=-\mathbf c^\dagger(\mathbf\Gamma+\mathbf\Gamma^\dagger)\mathbf c$, and the outgoing power is $\pmb \varepsilon_{\rm out}^\dagger\pmb \varepsilon_{\rm out}=\mathbf c^\dagger \mathbf D\mathbf D^\dagger\mathbf c$. The energy conservation constraint in this case leads to $\mathbf D\mathbf D^\dagger=\mathbf\Gamma+\mathbf\Gamma^\dagger$. For $\pmb \varepsilon_{\rm in}\neq0$ and at steady state where $d\mathbf c/dt=0$, we find $\mathbf D\mathbf K^{-1}\mathbf\Delta=\mathbf\Delta \mathbf K^{-1\dagger}\mathbf D^\dagger$ and $\mathbf \Gamma^\dagger\mathbf K^{-1\dagger}\mathbf D^\dagger+\mathbf D\mathbf K^{-1}\mathbf \Gamma=\mathbf \Gamma+\mathbf \Gamma^\dagger$, valid for any matrix $\mathbf\Delta$ \cite{suh2004temporal,zhao2019connection}. Thus we can determine that $\mathbf D=\mathbf K$, and
\begin{equation}
	\mathbf\Gamma=\left(\begin{array}{cc}
		\Gamma_a & \Gamma_{b\rightarrow a} \\
		\Gamma_{a\rightarrow b} & \Gamma_b
	\end{array}
	\right)=\left(\begin{array}{cc}
		\frac{|k_{a1}|^2+|k_{a2}|^2}{2} & k_{a2}k_{b2}^*e^{i\phi} \\
		k_{a1}^*k_{b1}e^{i\phi} & \frac{|k_{b1}|^2+|k_{b2}|^2}{2}
	\end{array}
	\right).
\end{equation} 
Here, $\Gamma_a$, $\Gamma_b$ are the external damping rates into the waveguide modes, while $\Gamma_{a\rightarrow b(b\rightarrow a)}$ is the radiative interaction between the two resonators via waveguide.

\section{Transmission properties of the model}
\label{trans}

We now set out to show how to derive the transmission parameters for the forward propagation ($\varepsilon_{1\rightarrow}$ drive) and the backward propagation ($\varepsilon_{2\leftarrow}$ drive) from the basic equations in Section \ref{MB}. In the long-time limit, the operators in the Hamiltonian [Eq. (\ref{ha1})] reduce to their expectation values. For conciseness, we omit the notation $\langle.\rangle$, and the terms $a$, $b$ and $\varepsilon_{\ell\rightleftarrows}$ now denote the complex amplitudes of the resonant and waveguide modes, respectively. When achieving the steady state with $\dot a=0$, $\dot b=0$, Eq. (\ref{tcmt11}) results in the nonlinear response of the system to the counter-propagating drives,
\begin{equation}
	\begin{aligned}
		D(x)a&=\varepsilon_{1\rightarrow}[-k_{a1}(i\Delta_b+\gamma_b+\Gamma_b)+k_{b1}e^{i\phi}(\Gamma_{b\rightarrow a}+iJ)]\\
		&\!\!\!\!\!\!+\varepsilon_{2\leftarrow}[-k_{a2}e^{i\phi}(i\Delta_b+\gamma_b+\Gamma_b)+k_{b2}(\Gamma_{b\rightarrow a}+iJ)],
	\end{aligned}
	\label{a1}
\end{equation} 
\begin{equation}
	\begin{aligned}
		D(x)b&\\
		&\!\!\!\!\!\!\!\!\!\!\!\!\!\!\!\!\!\!=\varepsilon_{1\rightarrow}[k_{a1}(\Gamma_{a\rightarrow b}+iJ^*)
		-k_{b1}e^{i\phi}(i\Delta_a+\gamma_a+\Gamma_a+2ix)]\\
		&\!\!\!\!\!\!\!\!\!\!\!\!\!\!\!\!\!\!+\varepsilon_{2\leftarrow}[k_{a2}e^{i\phi}(\Gamma_{a\rightarrow b}+iJ^*)-k_{b2}(i\Delta_a+\gamma_a+\Gamma_a+2ix)],
	\end{aligned}
	\label{b1}
\end{equation} 
where 
\begin{equation}
	\begin{aligned}
	D(x)&=(\Gamma_{a\rightarrow b}+iJ^*)(\Gamma_{b\rightarrow a}+iJ)\\
	&\!\!\!\!\!\!-(i\Delta_a+\gamma_a+\Gamma_a+2ix)(i\Delta_b+\gamma_b+\Gamma_b),
	\end{aligned}
\end{equation}
and $x=U|a|^2$. The value of $x$ is obtained by solving the cubic equation that results from taking the square of the norm of both sides of Eq. (\ref{a1}). This may yield multiple solutions for $x$, which will be discussed in detail in Section \ref{nonlinear}.

The transmission parameters ($t_\rightarrow$ and $t_\leftarrow$) for both forward and backward driving direction ($\varepsilon_{1\rightarrow}$ and $\varepsilon_{2\leftarrow}$ drive, respectively) can be determined analytically using the steady-state solutions Eq. (\ref{a1}), (\ref{b1}), and the input-output relation Eq. (\ref{tcmt12}). When we set the backward drive $\varepsilon_{2\leftarrow}=0$, $x=x_\rightarrow$ is determined, the transmission parameter for the forward (rightward) propagation from port $1$ is
\begin{equation}
	\begin{aligned}
		t_\rightarrow=\frac{\varepsilon_{2\rightarrow}}{\varepsilon_{1\rightarrow}}&=\frac{e^{i\phi}}{D(x_\rightarrow)}[iJ^*(\Gamma_{b\rightarrow a}-\Gamma^*_{a\rightarrow b}+iJ)\\
		&\!\!\!\!\!\!\!\!\!\!\!\!\!\!\!\!\!\!\!\!\!\!\!\!\!\!\!\!\!\!-(i\Delta_a+\gamma_a-\frac{|k_{a1}|^2-|k_{a2}|^2}{2}+2ix_\rightarrow)\\
		&\!\!\!\!\!\!\!\!\!\!\!\!\!\!\!\!\!\!\!\!\!\!\!\!\!\!\!\!\!\!\times(i\Delta_b+\gamma_b-\frac{|k_{b1}|^2-|k_{b2}|^2}{2})],
	\end{aligned}
	\label{tab}
\end{equation} 
and when we set the forward driving $\varepsilon_{1\rightarrow}=0$, $x=x_\leftarrow$ is determined, the transmission parameter for the backward (leftward) propagation from port $2$ is 
\begin{equation}
	\begin{aligned}
		t_\leftarrow=\frac{\varepsilon_{1\leftarrow}}{\varepsilon_{2\leftarrow}}&=\frac{e^{i\phi}}{D(x_\leftarrow)}[iJ(\Gamma_{a\rightarrow b}-\Gamma^*_{b\rightarrow a}+iJ^*)\\
		&\!\!\!\!\!\!\!\!\!\!\!\!\!\!\!\!\!\!\!\!\!\!\!\!\!\!\!\!\!\!-(i\Delta_a+\gamma_a+\frac{|k_{a1}|^2-|k_{a2}|^2}{2}+2ix_\leftarrow)\\
		&\!\!\!\!\!\!\!\!\!\!\!\!\!\!\!\!\!\!\!\!\!\!\!\!\!\!\!\!\!\!\times(i\Delta_b+\gamma_b+\frac{|k_{b1}|^2-|k_{b2}|^2}{2})].
	\end{aligned}
	\label{tba}
\end{equation} 
The parameter $x$ in the transmission parameter contains the effect of nonlinearity.

When considering just one nonlinear resonator, $a$, with $\phi=J=k_{b1}=k_{b2}=0$, the transmission parameters for both propagation directions are reduced to
\begin{equation}
	t_{\rightarrow}=\frac{i(\Delta_a+2x_\rightarrow)+\gamma_a-(|k_{a1}|^2-|k_{a2}|^2)/2}{i(\Delta_a+2x_\rightarrow)+\gamma_a+\Gamma_a},
\end{equation} 
\begin{equation}
	t_{\leftarrow}=\frac{i(\Delta_a+2x_\leftarrow)+\gamma_a+(|k_{a1}|^2-|k_{a2}|^2)/2}{i(\Delta_a+2x_\leftarrow)+\gamma_a+\Gamma_a}.
\end{equation} 
The nonlinear response of the resonator $a$ to the input fields is
\begin{equation}
	a=\frac{k_{a1}\varepsilon_{1\rightarrow}+k_{a2}\varepsilon_{2\leftarrow}}{i(\Delta_a+2x)+\gamma_a+\Gamma_a}.
	\label{sra}
\end{equation} 
When $|k_{a1}| = |k_{a2}|$, as required by time reversal symmetry \cite{suh2004temporal,zhao2019connection}, the system exhibits reciprocal transmission even if $U\neq0$, as indicated in Eq. (\ref{sra}) where $x$ will be the same for counter-propagating drives $\varepsilon_{1\rightarrow}$ and $\varepsilon_{2\leftarrow}$. In other models \cite{cotrufo2021nonlinearity1}, a different direct scattering matrix $\mathbf C$ results in different constraints when time reversal symmetry is imposed. This leads to nonreciprocal behavior in systems with a single nonlinear resonator.

\section{Is Nonreciprocity without Nonlinearity Possible?}
\label{linear}

We first investigate the system without any nonlinear effects, setting $U=0$, thus $x=U|a|^2=0$. The presence of the second terms within parentheses in Eq. (\ref{tab}) and Eq. (\ref{tba}) indicates that even in the absence of a direct coupling $J$ between the resonators $a$ and $b$, nonreciprocity of the transmission can manifest if $|k_{a1}|\neq|k_{a2}|$ or $|k_{b1}|\neq|k_{b2}|$. Conversely, If $|k_{a1}|=|k_{a2}|$ and $|k_{b1}|=|k_{b2}|$, then the nonreciprocity condition hinges on the condition that the coupling constant $J\neq0$.

In the system with $|k_{a1}|=|k_{a2}|$ and $|k_{b1}|=|k_{b2}|$, from Eq. (\ref{tab}) and Eq. (\ref{tba}), we can find that the condition for nonreciprocal transmission ($t_\rightarrow\neq t_\leftarrow$) is
\begin{equation}
	\begin{aligned}
		{\rm Re}[J(\Gamma_{a\rightarrow b}-\Gamma^*_{b\rightarrow a})]\neq0.\\
	\end{aligned}
	\label{gc}
\end{equation} 

For this system equipped with a real coupling constant $J$, the condition in Eq. (\ref{gc}) is reduced to
\begin{equation}
	{\rm Re}(\Gamma_{a\rightarrow b})\neq{\rm Re}(\Gamma_{b\rightarrow a}),
	\label{tr1}
\end{equation} 
which is determined by the phases of the coupling constants $k_{a\ell}$ and $k_{b\ell}$, along with the phase shift $\phi$ accumulated between the two resonators. When $k_{a\ell}$ and $k_{b\ell}$ are real, if $\cos\phi=0$, then there is no nonreciprocity; if $\cos\phi\neq0$, then the nonreciprocity condition given by Eq. (\ref{tr1}) simplifies to the requirement that $k_{a1}k_{b1} \neq k_{a2}k_{b2}$. Since $|k_{a1}|=|k_{a2}|$ and $|k_{b1}|=|k_{b2}|$, achieving nonreciprocity requires $k_{a1}=-k_{a2}$ and $k_{b1}=k_{b2}$, or $k_{b1}=-k_{b2}$ and $k_{a1}=k_{a2}$ with $\phi\neq (2N+1)\pi/2$ ($N\in\mathbb{Z}$). These are exactly the conditions under which experiments in \cite{wang2019nonreciprocity} are done.

In order to see how to achieve $k_{a1}=-k_{a2}$ or $k_{b1}=-k_{b2}$, let us conside one of the resonators to be magnetic. For a magnetic resonator, the coupling strength can flip its sign due to the reversal of the magnetic field's direction with counter-propagating directions. Most simply, a monochromatic wave is governed by $\mathbf B=\mathbf k\times\mathbf E/\omega$, where $\mathbf k$ denotes the wave vector and $\omega$ denotes the wave frequency. The orientation of $\mathbf B$ depends upon the direction of $\mathbf k$. Accordingly, the potential energy of a magnetic dipole $\mathbf m$ in a magnetic field $\mathbf B$, given by $E=-\mathbf m\cdot\mathbf B$, determines the sign of the coupling strength between the magnetic resonator and the incoming waves. Thus, when one of the resonators is magnetic, and the other is optical, i.e., $k_{a1}=-k_{a2}$ and $k_{b1}=k_{b2}$, or $k_{a1}=k_{a2}$ and $k_{b1}=-k_{b2}$, with $\phi\neq (2N+1)\pi/2$, the nonreciprocity condition, as stated in Eq. (\ref{tr1}), is achieved \cite{wang2019nonreciprocity}.  

In the system composed of two optical resonators where the coupling strengths are equal for each resonator ($k_{a1}=k_{a2}$ and $k_{b1}=k_{b2}$), utilizing the phase of the complex coupling strength $J$ can aid in achieving nonreciprocal transmission. When $J$ is represented as $J=|J|e^{i\theta}$, the nonreciprocal transmission condition can be achieved even when Eq. (\ref{tr1}) is violated, i.e., ${\rm Re}(\Gamma_{a\rightarrow b})={\rm Re}(\Gamma_{b\rightarrow a})$. Specifically, setting $k_{a1}=k_{a2}=k_a$, $k_{b1}=k_{b2}=k_b$, Eq. (\ref{tab}) and Eq. (\ref{tba}) can be reduced to   
\begin{equation}
	t_\rightarrow=\frac{e^{i\phi}}{D(0)}[-2|J|k_{a}k_{b}e^{-i\theta}\sin\phi-|J|^2-(i\Delta_a+\gamma_a)(i\Delta_b+\gamma_b)],
\end{equation} 
\begin{equation}
	t_\leftarrow=\frac{e^{i\phi}}{D(0)}[-2|J|k_{a}k_{b}e^{i\theta}\sin\phi-|J|^2-(i\Delta_a+\gamma_a)(i\Delta_b+\gamma_b)],
\end{equation} 
which indicates that nonreciprocal transmission ($t_\rightarrow\neq t_\leftarrow$) can occur when $\sin\theta\sin\phi\neq0$, that is, $\theta\neq N\pi$, $\phi\neq N\pi$. A recent paper discusses nonreciprocity from complex $J$ within the context of magnetic films using the coherent Dzyaloshinskii-Moriya (DM) interaction \cite{zou2024dissipative}.

\section{Nonlinearity induced nonreciprocity if linearity does not allow nonreciprocity}
\label{nonlinear}

\begin{figure}[b]
	\includegraphics[width=9cm]{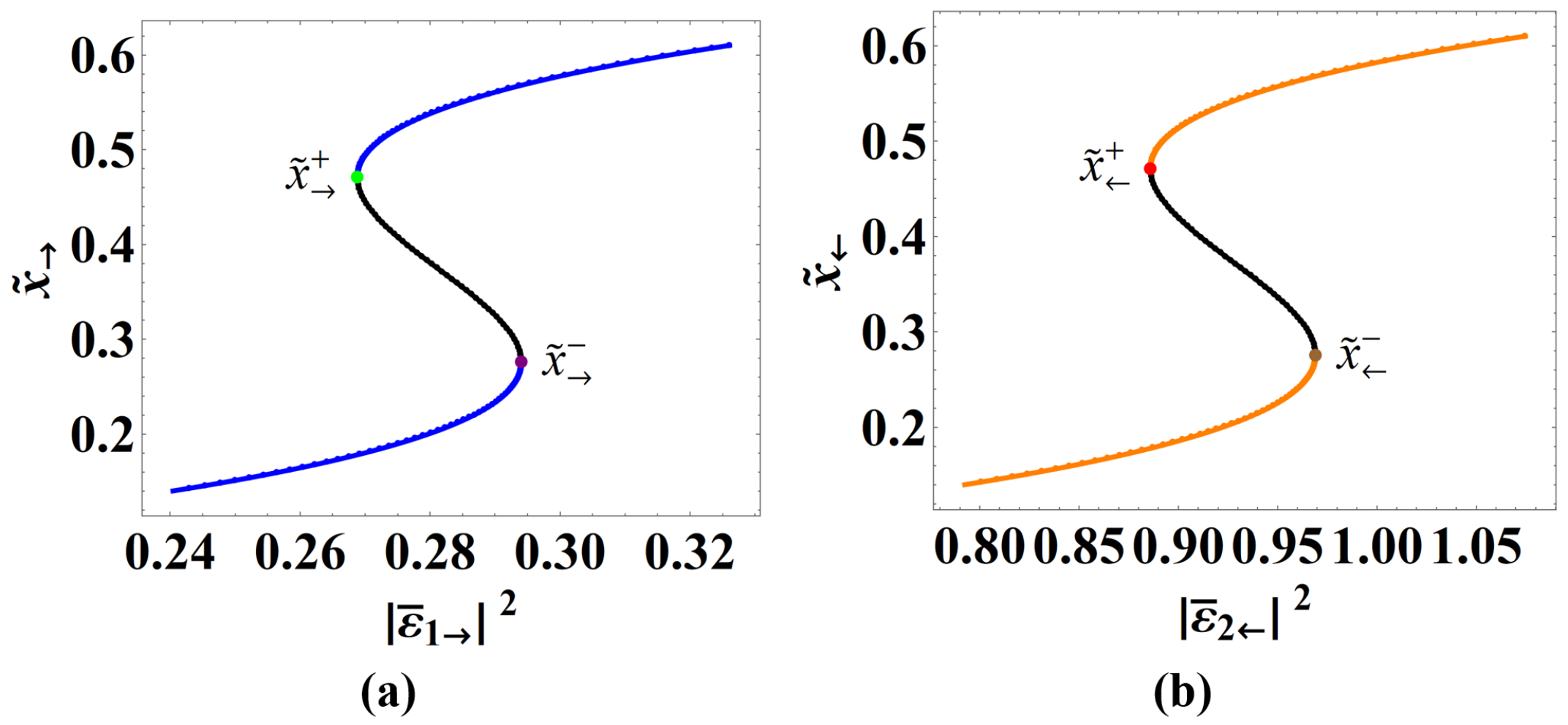}
	\caption{The $|\bar{\varepsilon}|^2$--$\tilde x$ curves, (a) for $\varepsilon_{1\rightarrow}$ drive, (b) for $\varepsilon_{2\leftarrow}$ drive. The turning points $\tilde x^\pm$ are as denoted in the figure. The black lines are for the unstable states. $\Gamma/2\pi=10$ MHz, $\tilde\gamma_a=\tilde\gamma_b=0.1$, $\tilde\Delta=-0.5$, $\phi=16\pi/15$.}
	\label{bis}
\end{figure}
\begin{figure}[b]
	\includegraphics[width=8.6cm]{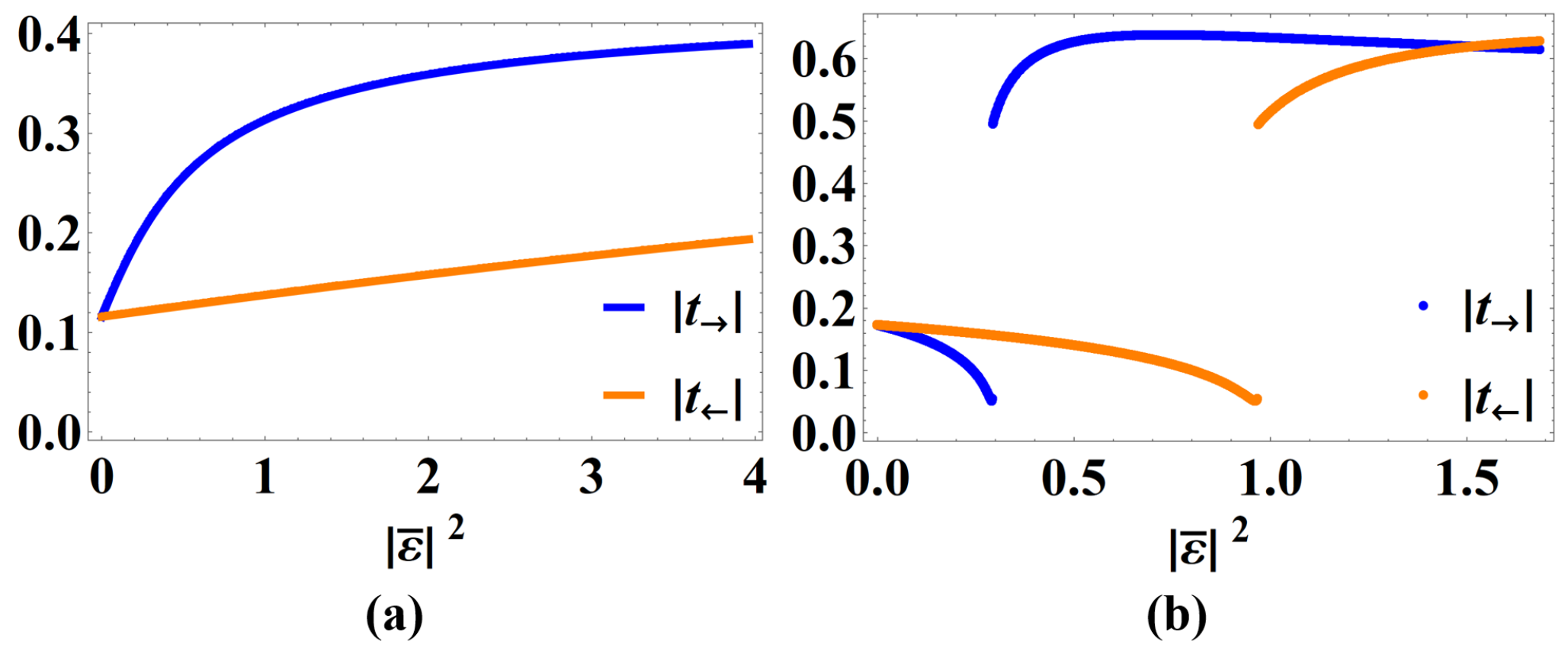}
	\caption{Transmission spectrum as a function of the scaled photon flux at the input $|\bar{\varepsilon}|^2$. The blue lines are for $\varepsilon_{1\rightarrow}$ drive, and the yellow lines are for $\varepsilon_{2\leftarrow}$ drive. $\tilde\gamma_a=\tilde\gamma_b=0.1$. (a) $\tilde\Delta=0.5$, $\phi=\pi/2$. (b) $\tilde\Delta=-0.5$, $\phi=16\pi/15$, consistent with Fig. \ref{bis}. Sudden jumps in $|t|$ at $\tilde x^-$ are found as drives increase.}
	\label{tvsp}
\end{figure}
\begin{figure}[b]
	\includegraphics[width=8.6cm]{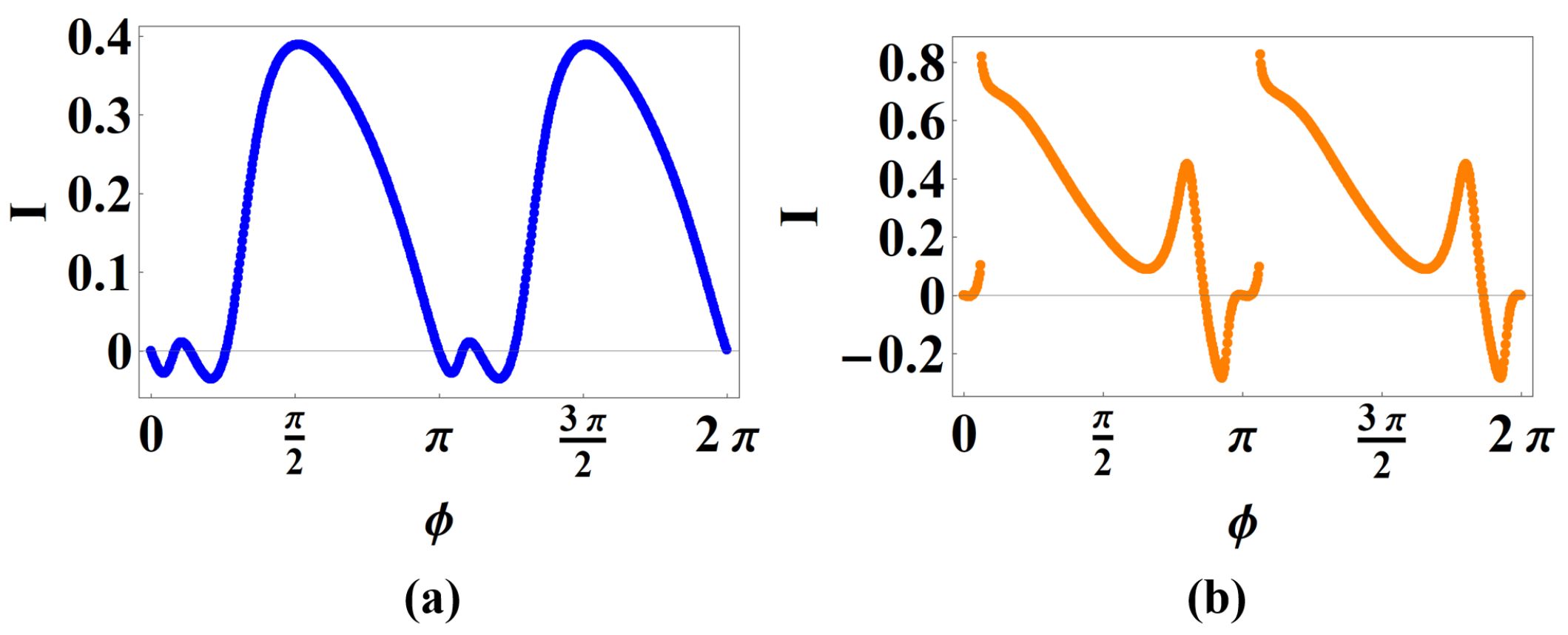}
	\caption{Nonreciprocity parameter $I=(|t_\rightarrow|-|t_\leftarrow|)/(|t_\rightarrow|+|t_\leftarrow|)$ as a function of the phase shift $\phi$. The parameters are the same as Fig. \ref{tvsp} ($\tilde U>0$). (a) $|\bar{\varepsilon}|^2=2$, $\tilde\Delta=0.5$. (b) $|\bar{\varepsilon}|^2=0.9$, $\tilde\Delta=-0.5$.}
	\label{eff}
\end{figure}
\begin{figure}[b]
	\includegraphics[width=8.6cm]{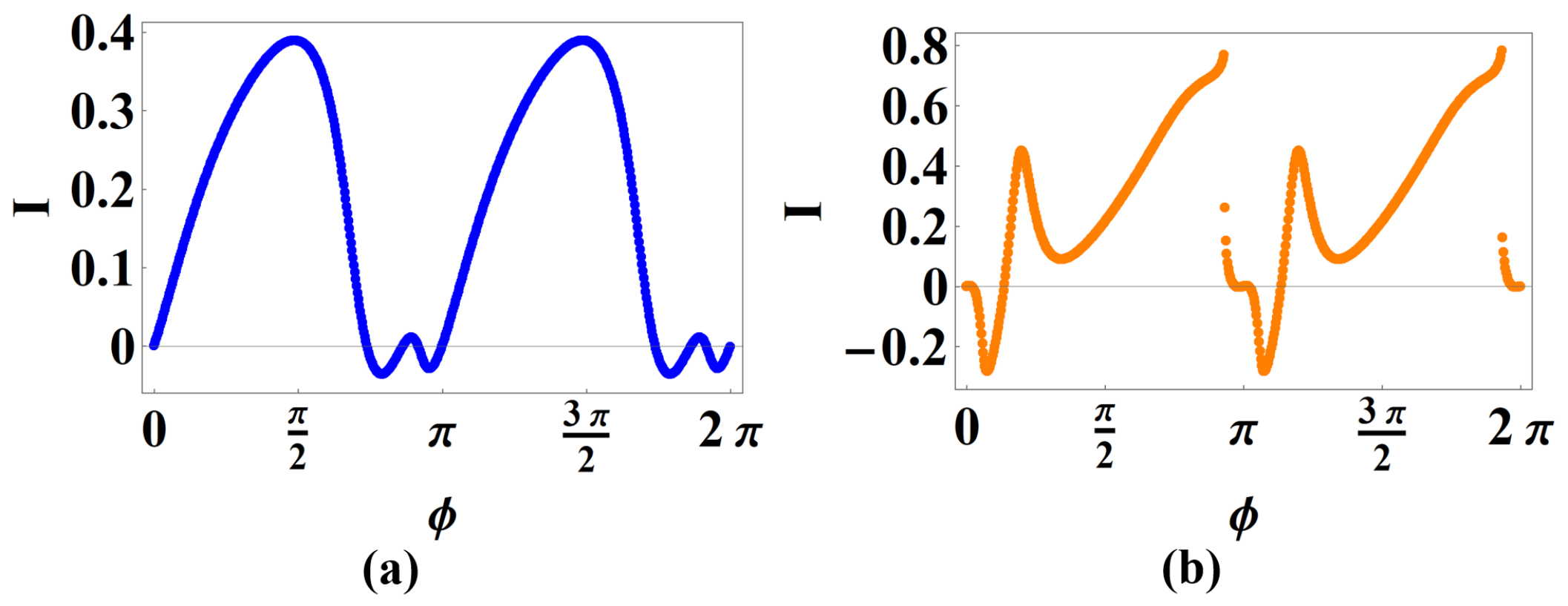}
	\caption{Nonreciprocity parameter $I$ as a function of the phase shift $\phi$. Now we consider the case when $\tilde U<0$. (a) $|\bar{\varepsilon}|^2=2$, $\tilde\Delta=-0.5$. (b) $|\bar{\varepsilon}|^2=0.9$, $\tilde\Delta=0.5$.}
	\label{eff_U}
\end{figure}
As discussed in Section \ref{linear}, if the direct coupling $J$ is real and all coupling constants are identical, then the linear medium fails to achieve nonreciprocity. Here, we show how nonlinearity can produce significant nonreciprocal behavior. We note that the nonlinearity induced nonreciprocity has been studied before. Examples include intrinsic nonlinearities of two level systems \cite{hamann2018nonreciprocity,roy2010few}; nonlinearities of Brillouin media \cite{dong2015brillouin,kim2015non}; nonlinear PT symmetric media \cite{peng2014parity,liu2014regularization,chang2014parity}; systems with a different scattering matrix subjected to conditions imposed by time reversal symmetry \cite{cotrufo2021nonlinearity1,cotrufo2021nonlinearity2}. We now consider a Kerr nonlinear system where there is no coherent coupling $J$ between the two resonators, i.e., $J=0$, and the coupling constants $k_{a\ell}$, $k_{b\ell}$ are real and identical for counter-propagating directions, i.e., $k_{a1}=k_{a2}$, $k_{b1}=k_{b2}$. However, the mode denoted by $a$ is now characterized by third-order nonlinearity, i.e., $U\neq0$. For simplification, we also assume that $\Delta_a=\Delta_b=\Delta$, $k_{a\ell}=k_{b\ell}=\sqrt{\Gamma}$, where $\Gamma$ denotes the total radiative decay rate of an individual resonator into the waveguide port. At the steady state with $\dot a=0$, $\dot b=0$, we have two equations for $a$ and $b$ followed from Eq. (\ref{tcmt11}) and Eq. (\ref{tcmt12}) ($a$, $b$ and $\varepsilon_{\ell\rightleftarrows}$ denote the complex amplitudes as discussed in Section \ref{trans}), 
\begin{equation}
	\begin{aligned}
	2i\tilde U|a|^2a+(i\tilde \Delta+\tilde \gamma_a+1)a+ e^{i\phi}b=\tilde \varepsilon_{1\rightarrow}+e^{i\phi}\tilde \varepsilon_{2\leftarrow},\\
	(i\tilde \Delta+\tilde \gamma_b+1)b+e^{i\phi}a=e^{i\phi}\tilde \varepsilon_{1\rightarrow}+\tilde \varepsilon_{2\leftarrow},
	\end{aligned}
	\label{nle}
\end{equation}
along with two equations for the input-output relation,
\begin{equation}
	\begin{aligned}
	\tilde \varepsilon_{2\rightarrow}=e^{i\phi}\tilde \varepsilon_{1\rightarrow}- e^{i\phi} a- b,\\
	\tilde \varepsilon_{1\leftarrow}=e^{i\phi}\tilde \varepsilon_{2\leftarrow}- a- e^{i\phi}b,
	\end{aligned}
	\label{nlio}
\end{equation}
all of which are normalized with the scaling factor $\Gamma$, leading to the definitions: $\tilde U=U/\Gamma$, $\tilde\Delta=\Delta/\Gamma$, $\tilde\gamma_a=\gamma_a/\Gamma$, $\tilde\gamma_b=\gamma_b/\Gamma$, $\tilde \varepsilon_{\ell\rightleftarrows}=\varepsilon_{\ell\rightleftarrows}/\sqrt\Gamma$. From Eq. (\ref{nle}), the nonlinear response of resonator $a$ is obtained, 
\begin{equation}
	\begin{aligned}
	4\tilde x^3+4({\rm Re}\ \tilde\delta)\tilde x^2+|\tilde\delta|^2\tilde x&\\
	&\!\!\!\!\!\!\!\!\!\!\!\!\!\!\!\!\!\!\!\!\!\!\!\!\!\!\!\!\!\!\!\!\!\!\!\!\!\!\!\!\!\!=\pm\frac{|(i\tilde\Delta+\tilde\gamma_b+1-e^{2i\phi})\bar \varepsilon_{1\rightarrow}+(i\tilde\Delta+\tilde\gamma_b) e^{i\phi}\bar \varepsilon_{2\leftarrow}|^2}{\tilde\Delta^2+(\tilde\gamma_b+1)^2},
	\end{aligned}
	\label{nlx}
\end{equation}
where we introduce $\tilde x=\tilde U|a|^2$, $i\tilde\delta=i\tilde \Delta+\tilde\gamma_a+1-\frac{e^{2i\phi}}{i\tilde \Delta+\tilde\gamma_b+1}$, $\bar \varepsilon_{1\rightarrow}=\sqrt{|\tilde U|}\tilde \varepsilon_{1\rightarrow}$, $\bar \varepsilon_{2\leftarrow}=\sqrt{|\tilde U|}\tilde \varepsilon_{2\leftarrow}$, with a plus sign in right hand side for $\tilde U>0$, and a minus sign for $\tilde U<0$. In order to achieve nonreciprocity, in Eq. (\ref{nlx}), the coefficients before the input terms $\bar \varepsilon_{1\rightarrow}$ and $\bar \varepsilon_{2\leftarrow}$ should differ, to ensure distinct nonlinear responses to the counter-propagating drives. Nonreciprocity for the Kerr nonlinear system then hinges on the condition $e^{2i\phi}\neq1$, implying 
\begin{equation}
	\phi\neq N\pi, \qquad(N\in\mathbb{Z}).
	\label{lnc}
\end{equation}

Note that the cubic equation [Eq. (\ref{nlx})] can have three real roots under the conditions
\begin{equation}
	({\rm Re}\ \tilde\delta)^2>3({\rm Im}\ \tilde\delta)^2,\qquad \tilde U{\rm Re}\ \tilde\delta<0,
	\label{c}
\end{equation}
leading to a bistability response \cite{miao2023engineering}. Now we assmue $\tilde U>0$. We plot $\tilde x$ against the scaled photon flux at the input $|\bar{\varepsilon}|^2$ with experimentally feasible parameters which satisfies Eq. (\ref{c}), illustrated in Fig. \ref{bis}. The $|\bar{\varepsilon}|^2$--$\tilde x$ curve features two turning points subject to $d(|\bar \varepsilon|^2)/d\tilde x=0$. With specific region of the amplitude of the input, there exist three real solutions for $\tilde x$, with two representing stable states and one an unstable state. Fulfilling the nonreciprocity criterion in Eq. (\ref{lnc}) results in two distinct bistability regions for counter-propagating drives, as illustrated in Fig. \ref{bis}(a) and Fig. \ref{bis}(b). As drives increase nearing the turning points $\tilde x^-$, sharp jumps in $\tilde x$ from one stable state to another occur.

The transmission parameter can be obtained by substituting the expressions for $a$ and $b$, derived from Eq. (\ref{nle}), into the input-output relation given by Eq. (\ref{nlio}). Once the value of $\tilde x$ is determined from Eq. (\ref{nlx}) for incoming drives ($\varepsilon_{1 \rightarrow}$, $\varepsilon_{2 \leftarrow}$) from each direction, we can calculate out the value of $a$ from Eq. (\ref{nle}),
\begin{equation}
	a=\frac{(i\tilde\Delta+\tilde\gamma_b+1-e^{2i\phi})\tilde \varepsilon_{1\rightarrow}+(i\tilde\Delta+\tilde\gamma_b) e^{i\phi}\tilde \varepsilon_{2\leftarrow}}{(i\tilde\Delta+\tilde\gamma_a+1+2i\tilde x)(i\tilde\Delta+\tilde\gamma_b+1)-e^{2i\phi}},
	\label{aa}
\end{equation}
which exhibits distinct behaviours based on the propagation direction if Eq. (\ref{lnc}) is satisfied. When we set the backward driving $\varepsilon_{2\leftarrow}=0$, the steady-state solution $\tilde x=\tilde x_{\rightarrow}$ is determined from Eq. (\ref{nlx}), the transmission parameter for the forward (rightward) direction is
\begin{equation}
	t_\rightarrow=\frac{\varepsilon_{2\rightarrow}}{\varepsilon_{1\rightarrow}}=\frac{e^{i\phi}(i\tilde\Delta+\tilde\gamma_a+2i\tilde x_\rightarrow)(i\tilde\Delta+\tilde\gamma_b) }{(i\tilde\Delta+\tilde\gamma_a+1+2i\tilde x_\rightarrow)(i\tilde\Delta+\tilde\gamma_b+1)-e^{2i\phi}}.
\end{equation}
Similarly, when we set the forward driving $\varepsilon_{1\rightarrow}=0$, the steady-state solution $\tilde x=\tilde x_{\leftarrow}$ is determined from Eq. (\ref{nlx}), the transmission parameter for the backward (leftward) direction is
\begin{equation}
	t_\leftarrow=\frac{\varepsilon_{1\leftarrow}}{\varepsilon_{2\leftarrow}}=\frac{e^{i\phi}(i\tilde\Delta+\gamma_a+2i\tilde x_\leftarrow)(i\tilde\Delta+\tilde\gamma_b) }{(i\tilde\Delta+\tilde\gamma_a+1+2i\tilde x_\leftarrow)(i\tilde\Delta+\tilde\gamma_b+1)-e^{2i\phi}}.
\end{equation}

Both transmission parameters share the same dependence on the parameter $\tilde x$, which can be different between counter-propagating drives. When the phase shift $\phi= N\pi$ (violating Eq. (\ref{lnc}) and leading to $\tilde x_\rightarrow= \tilde x_\leftarrow$), the system exhibits reciprocal behavior, i.e., $t_\rightarrow=t_\leftarrow$. Nonetheless, with $\phi\neq N\pi$, resulting in $\tilde x_\rightarrow\neq \tilde x_\leftarrow$, it leads to $t_\rightarrow\neq t_\leftarrow$, indicating nonreciprocal transmission. The drive power $P$ is related to the coherent photon flux at the input, through $P=\hbar\omega_d|\varepsilon|^2$, for the respective directions. Eq. (\ref{nlx}) shows that the value of $\tilde x$ depends on the scaled input, i.e., $|\bar\epsilon|^2=UP/\hbar\omega_d\Gamma^2$. For a certain value of $\tilde x$, the input power $P$ will be lower if $U$ becomes higher.

The nonreciprocal behavior of the system is illustrated in Fig. \ref{tvsp}, where the transmission amplitude $|t|$ is plotted against the scaled photon flux at the input $|\bar\epsilon|^2$ for counter-propagating inputs. In Fig. \ref{tvsp}(a), the phase $\phi=\pi/2$ satisfies the nonreciprocity condition in Eq. (\ref{lnc}), but does not fulfill the bistability condition in Eq. (\ref{c}). As a comparison, in Fig. \ref{tvsp}(b) with the same parameters as Fig. \ref{bis}, the phase $\phi=16\pi/15$, slightly detuned from $\pi$, satisfies both the nonreciprocity and bistability conditions. As counter-propagating drives increases nearing $\tilde x^-$, corresponding sudden jumps in the transmission amplitude $|t|$ within distinct regions found in Fig. \ref{tvsp}(b) due to bistability aids in achieving large nonreciprocity. 
To measure the nonreciprocity, we define the nonreciprocity parameter of the nonreciprocal system as $I=(|t_\rightarrow|-|t_\leftarrow|)/(|t_\rightarrow|+|t_\leftarrow|)$. This is graphically represented as a function of the phase shift $\phi$, with a period of $\pi$, as depicted in Fig. \ref{eff}. Additionally, we plot the nonreciprocity parameter when $U<0$ as shown in Fig. \ref{eff_U}.

In this section, both the resonators are considered optical. When the resonator $a$ is switched to magnetic, i.e., $k_{a1}=-k_{a2}$ and $k_{b1}=k_{b2}$, it results in merely a phase transformation, $\phi\rightarrow\phi+\pi/2$, as discussed in Section \ref{appen}.

\section{Nonreciprocity in excitation of each resonator}
\label{nex}

\begin{figure}[b]
	\includegraphics[width=8.6cm]{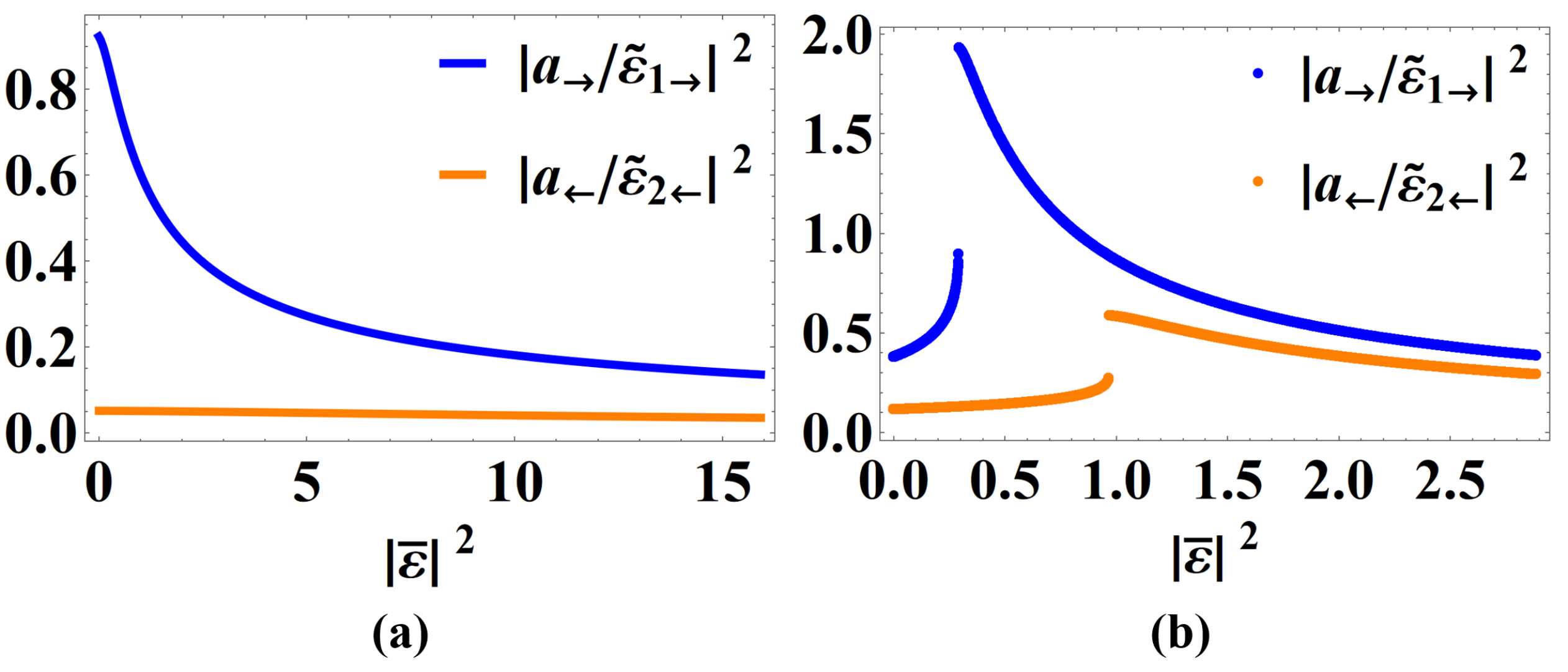}
	\caption{The ratio of the excitation energy in nonlinear resonator $a$ ($U>0$) relative to the input photon flux for counter-propagating drives as a function of drive power. (a) $\tilde\Delta=0.5$, $\phi=\pi/2$. (b) $\tilde\Delta=-0.5$, $\phi=16\pi/15$.}
	\label{excitation}
\end{figure}
Driving the two resonators coherently from either the left-side or the right-side give rise to the excitation energies in the resonators. For magnetic resonators, the resulting excited magnetization can be electrically detected through the spin currents via the inverse spin Hall effect (ISHE) \cite{bai2017cavity}. For the linear system ($U=0$), when driving from the left-side, i.e., $\varepsilon_{1\rightarrow}\neq0$ and $\varepsilon_{2\leftarrow}=0$, the energy ratio of the resonant mode $a$ relative to the incoming power is found from Eq. (\ref{a1}) and Eq. (\ref{b1}),
\begin{equation}
	\begin{aligned}
		\left|\frac{a_\rightarrow}{\varepsilon_{1\rightarrow}}\right|^2&\\
		&\!\!\!\!\!\!\!\!\!\!\!\!=\frac{1}{|D(0)|^2}(|k_{a1}|^2|i\Delta_b+\gamma_b+\Gamma_b|^2+|k_{b1}|^2|\Gamma_{b\rightarrow a}+iJ|^2)\\
		&\!\!\!\!\!\!\!\!\!\!\!\!-\frac{1}{|D(0)|^2}[\Gamma_{a\rightarrow b}(-i\Delta_b+\gamma_b+\Gamma_b)(\Gamma_{b\rightarrow a}+iJ)+{\rm c.\ c.}].
	\end{aligned}
\end{equation} 
When driving from the right-side, i.e., $\varepsilon_{2\leftarrow}\neq0$ and $\varepsilon_{1\rightarrow}=0$, the energy ratio of the resonant mode $a$ relative to the incoming power is similarly found as
\begin{equation}
	\begin{aligned}
		\left|\frac{a_\leftarrow}{\varepsilon_{2\leftarrow}}\right|^2&\\
		&\!\!\!\!\!\!\!\!\!\!\!\!=\frac{1}{|D(0)|^2}(|k_{a2}|^2|i\Delta_b+\gamma_b+\Gamma_b|^2+|k_{b2}|^2|\Gamma_{b\rightarrow a}+iJ|^2)\\
		&\!\!\!\!\!\!\!\!\!\!\!\!-\frac{1}{|D(0)|^2}[\Gamma_{b\rightarrow a}^*(-i\Delta_b+\gamma_b+\Gamma_b)(\Gamma_{b\rightarrow a}+iJ)+{\rm c.\ c.}].
	\end{aligned}
\end{equation} 

Nonreciprocal excitations can be achieved by ensuring asymmetric external emission into counter-propagating waveguide modes, characterized by differing coupling strengths, i.e., $|k_{a1}|\neq|k_{a2}|$ or $|k_{b1}|\neq|k_{b2}|$, similar to nonreciprocal transmission as discussed in Section \ref{linear}. 

For a system where $|k_{a1}|=|k_{a2}|$ and $|k_{b1}|=|k_{b2}|$, the necessary conditions to achieve nonreciprocal excitations ($\left|{a_\rightarrow}/{\varepsilon_{1\rightarrow}}\right|^2\neq\left|{a_\leftarrow}/{\varepsilon_{2\leftarrow}}\right|^2$) are
\begin{equation}
	\Gamma_{a\rightarrow b}-\Gamma_{b\rightarrow a}^*\neq0,\qquad \Gamma_{b\rightarrow a}+iJ\neq0.
	\label{ex}
\end{equation}  
When $k_{a\ell}$ and $k_{b\ell}$ are real, the first condition in Eq. (\ref{ex}) is reduced to $\phi\neq N\pi$. This condition demonstrates that the reciprocity of excitations can be broken through the manipulation of the phase shift accumulated between two resonators.

For the Kerr nonlinear system with $\Delta_a=\Delta_b=\Delta$, $k_{a\ell}=k_{b\ell}=\sqrt{\Gamma}$, and $J=0$, the necessary condition for nonreciprocal exitations is $\phi\neq N\pi$. This condition, serving as the basis for nonreciprocal transmission as discussed in Section \ref{nonlinear}, can similarly be extracted from Eq. (\ref{nlx}). We plot the ratio ($|a/\tilde \varepsilon_{1\rightarrow}|^2$, $|a/\tilde \varepsilon_{2\leftarrow}|^2$) for counter-propagating drives as a function of the scaled photon flux as shown in Fig. \ref{excitation}.

Our Kerr nonlinear system can be configured with the resonator $a$ as either optical type or magnetic type for experimental realization. For an optical resonator, anharmonicity originates from nonlinear response of the electrical polarization. Recent experimental advancements have showed cavities with significant Kerr nonlinearity, measured by a Kerr coefficient of $U/2\pi=-12.2\pm0.1$ kHz/Photon \cite{zoepfl2023kerr}. For a magnetic resonator, anharmonicity originates from the nonlinear magnetization. Recent research into nonlinearities within ferrimagnetic spheres indicate that ferromagnetic materials, such as Yttrium Iron Garnet (YIG), can exhibit robust coupling with microwave fields against temperature. With the growing research interest in YIG, it stands out as an active Kerr medium for exploring nonlinearity induced nonreciprocity \cite{wang2016magnon,wang2018bistability}.

\section{Nonreciprocity Induced by Nonlinearity with a Magnetic Resonator}
\label{appen}

In this Section, We investigate the transmission property of the Kerr nonlinear system, focusing on a case different from the one presented in Section \ref{nonlinear}. Here, the resonator $a$ is magnetic with third-order nonlinearity, while the resonator $b$ is optical, characterized by different conditions $k_{a1}=-k_{a2}$ and $k_{b1}=k_{b2}$. Similarly, there is no coherent coupling $J$ between the two resonators ($J=0$), the coupling constants $k_{a\ell}$, $k_{b\ell}$ are considered real. For convenience, we also assume $\Delta_a=\Delta_b=\Delta$, $k_{a1}=-k_{a2}=\sqrt{\Gamma}$, $k_{b\ell}=\sqrt{\Gamma}$. At the steady state with $\dot a=0$, $\dot b=0$,
\begin{equation}
	\begin{aligned}
		2i\tilde U|a|^2a+(i\tilde \Delta+\tilde \gamma_a+1)a- e^{i\phi}b=\tilde \varepsilon_{1\rightarrow}-e^{i\phi}\tilde \varepsilon_{2\leftarrow},\\
		(i\tilde \Delta+\tilde \gamma_b+1)b+e^{i\phi}a=e^{i\phi}\tilde \varepsilon_{1\rightarrow}+\tilde \varepsilon_{2\leftarrow},
	\end{aligned}
	\label{nle2}
\end{equation}
along with two equations for the input-output relation,
\begin{equation}
	\begin{aligned}
		\tilde \varepsilon_{2\rightarrow}=e^{i\phi}\tilde \varepsilon_{1\rightarrow}- e^{i\phi} a- b,\\
		\tilde \varepsilon_{1\leftarrow}=e^{i\phi}\tilde \varepsilon_{2\leftarrow}+ a- e^{i\phi}b,
	\end{aligned}
	\label{nlio2}
\end{equation}
where $\tilde U$, $\tilde\Delta$, $\tilde\gamma_a$, $\tilde\gamma_b$, $\tilde \varepsilon_{\ell\rightleftarrows}$ are as defined in Section \ref{nonlinear}. From Eq. (\ref{nle2}), the nonlinear response of resonator $a$ is obtained, 
\begin{equation}
	\begin{aligned}
		4\tilde x^3+4({\rm Re}\ \tilde\delta')\tilde x^2+|\tilde\delta'|^2\tilde x&\\
		&\!\!\!\!\!\!\!\!\!\!\!\!\!\!\!\!\!\!\!\!\!\!\!\!\!\!\!\!\!\!\!\!\!\!\!\!\!\!\!\!\!\!=\frac{|(i\tilde\Delta+\tilde\gamma_b+1+e^{2i\phi})\bar \varepsilon_{1\rightarrow}-(i\tilde\Delta+\tilde\gamma_b) e^{i\phi}\bar \varepsilon_{2\leftarrow}|^2}{\tilde\Delta^2+(\tilde\gamma_b+1)^2},
	\end{aligned}
	\label{nlx2}
\end{equation}
where we define $\tilde x$, $\bar \varepsilon_{1\rightarrow}$, $\bar \varepsilon_{2\leftarrow}$ as outlined in Section \ref{nonlinear}, and introduce $i\tilde\delta'=i\tilde \Delta+\tilde\gamma_a+1+\frac{e^{2i\phi}}{i\tilde \Delta+\tilde\gamma_b+1}$, with the assumption that $\tilde U>0$. When we set the backward driving $\varepsilon_{2\leftarrow}=0$, the steady-state solution $\tilde x=\tilde x_{\rightarrow}$ is determined from Eq. (\ref{nlx2}), which then yields the transmission parameter for the forward (rightward) propagation,
\begin{equation}
	t_\rightarrow=\frac{\varepsilon_{2\rightarrow}}{\varepsilon_{1\rightarrow}}=\frac{e^{i\phi}(i\tilde\Delta+\tilde\gamma_a+2i\tilde x_\rightarrow)(i\tilde\Delta+\tilde\gamma_b) }{(i\tilde\Delta+\tilde\gamma_a+1+2i\tilde x_\rightarrow)(i\tilde\Delta+\tilde\gamma_b+1)+e^{2i\phi}}.
\end{equation}
Similarly, when we set the forward driving $\varepsilon_{1\rightarrow}=0$, the steady-state solution $\tilde x=\tilde x_{\leftarrow}$ is determined from Eq. (\ref{nlx2}), which then yields the transmission parameter for the backward (leftward) propagation,
\begin{equation}
	t_\leftarrow=\frac{\varepsilon_{1\leftarrow}}{\varepsilon_{2\leftarrow}}=\frac{e^{i\phi}(i\tilde\Delta+\gamma_a+2i\tilde x_\leftarrow)(i\tilde\Delta+\tilde\gamma_b) }{(i\tilde\Delta+\tilde\gamma_a+1+2i\tilde x_\leftarrow)(i\tilde\Delta+\tilde\gamma_b+1)+e^{2i\phi}}.
\end{equation}

Comparing the nonlinear response and the transmission parameters with those discussed in Section \ref{nonlinear}, the only crucial alteration is the transformation $e^{2i\phi} \rightarrow -e^{2i\phi}$, equivalent to a phase shift, $\phi \rightarrow \phi + \pi/2$. Consequently, in comparison to the case detailed in Section \ref{nonlinear}, switching the resonator $a$ from an optical to a magnetic type results in a transformation in phase, thus in this case with no coherent coupling ($J=0$), the system shows no nonreciprocity in the absence of nonlinearity but would show significant nonreciprocity if the phase is not equal to an odd multiple of $\pi/2$. With this transformation, the parameters in Fig. \ref{tvsp}(a) should be adjusted to $\phi=\pi$, and for Fig. \ref{tvsp}(b), the parameters should be adjusted to $\phi=47\pi/30$ to achieve corresponding results.

For the linear system ($U=0$) equipped with a complex coupling constant $J=|J|e^{i\theta}$, we consider the case when the coupling strengths are all real, the nonreciprocal transmission ($t_\rightarrow\neq t_\leftarrow$) can occur when $\cos\phi\cos\theta\neq0$, that is, $\theta\neq(2N+1)\pi/2$, $\phi\neq(2N+1)\pi/2$, which can be derived from Eq. (\ref{tab}) and Eq. (\ref{tba}). 

\section{Conclusions}

In our exploration of Kerr nonlinearity induced nonreciprocity in a system with two waveguide-coupled resonators, we unveil the potential of Kerr nonlinearity to manifest nonreciprocal behavior, particularly when linear systems fail to exhibit nonreciprocal behavior. We derive the conditions inducing nonreciprocity and study its dependency on various system parameters. Our analysis reveals that nonreciprocal transmission is not only possible but can be significantly enhanced through the strategic exploitation of nonlinearity, thus in the case with no coherent coupling $J=0$, the system shows no nonreciprocity in the absence of nonlinearity but can show significant nonreciprocity with an appropriately chosen phase $\phi$.

Finally, it may be noted that we can discuss nonreciprocity in a full quantum framework by converting the classical equation Eq. (\ref{tcmt11}) into nonlinear quantum Langevin equations for the mode operators. The very commonly used linearization treatment of quantum fluctuations \cite{Wallsbook} gives quantum fluctuations in terms of the values of the mean fields which are determined by Eq. (\ref{tcmt11}). Thus, the existence of nonreciprocal behavior of mean field equations would imply nonreciprocity in quantum fluctuations as well. We hope to report on these aspects in the future. 
\medskip

\textbf{Acknowledgements}

We thank the support of Air Force Office of Scientific Research (Award No FA-9550-20-1-0366) and the Robert A Welch Foundation (A-1943-20210327). G. S. A. thanks Prof. C. M. Hu and Y. Yang for the thesis of which enabled us to see all the experiments on nonreciprocity in magnetic systems.

\bibliography{main}

\end{document}